\documentclass[11pt]{article}

\usepackage{amsthm,amssymb,amsmath,latexsym}

\theoremstyle{plain}
\newtheorem{theorem}{Theorem}
\newtheorem{corollary}{Corollary}\newtheorem{lemma}{Lemma}

\theoremstyle{definition}
\newtheorem{definition}{Definition}

\numberwithin{equation}{section}

\newcommand{\mH}{\mathcal H}

\newcommand{\mX}{\mathcal X}
\newcommand{\mY}{\mathcal Y}

\newcommand{\mW}{\mathcal W}

\newcommand{\Listd}[2]{\textnormal{\textsf{(${#1}$,${#2}$)-List}}}
\newcommand{\List}[1]{\textnormal{\textsf{$({#1})$-List}}}

\newcommand{\Channel}[2]{$({#1},{#2})$-zero-error channel}

\newcommand{\cancel}[1]{}

\newcommand{\achieves}{\rightarrow}
\newcommand{\achievesF}{\stackrel{F}{\achieves}}

\newcommand{\nachievesF}{\stackrel{F}{\;\not\achieves\;}}

\begin{document}

\title{Zero-Error Communication over Networks}

\author{J\"urg Wullschleger\ \footnote{Work done while at the Comp.
    Sci. Dept., ETH Z\"urich, Switzerland}\\
    D\'epartement d'Informatique et R. O.\\
             Universit\'e de Montr\'eal, Canada\\
            e-mail: {\tt wullschj@iro.umontreal.ca}
}

\renewcommand{\today}{April 14, 2004}

\maketitle

\begin{abstract}
  Zero-Error communication investigates communication without any
  error. By defining channels without probabilities, results from
  Elias can be used to completely characterize which channel can
  simulate which other channels. We introduce the \emph{ambiguity} of
  a channel, which completely characterizes the possibility in
  principle of a channel to simulate any other channel.
 
  In the second part we will look at networks of players connected by
  channels, while some players may be corrupted. We will show how the
  ambiguity of a virtual channel connecting two arbitrary players can
  be calculated. This means that we can exactly specify what kind of
  zero-error communication is possible between two players in any 
  network of players connected by channels.
\end{abstract}

\section{Introduction}

The \emph{capacity} of a noisy channel was introduced by Shannon in
\cite{shanno48}. It defines the asymptotically maximal rate at which
bits can be transmitted by a channel with arbitrarily small error
probability.  Later in \cite{shanno56}, Shannon also proposed the
\emph{zero-error capacity} of a noisy channel, where not even an
arbitrarily small error is allowed in the transmission.  The small
change in the definition can cause a big difference in the value:
There are many channels for which the zero-error capacity is $0$,
whereas the ordinary capacity is positive\footnote{An example of such
  a channel is the binary-symmetric channel with error probability
  $\epsilon > 0$.}. Up to now, a formula for calculating the
zero-error capacity for any channel is still missing. In contrast to
the ordinary capacity, the zero-error capacity with feedback can be
bigger than the zero-error capacity without feedback.  The exact
formula for the zero-error capacity with feedback is known, and gives
an upper bound on the zero-error capacity without feedback.

Elias showed in \cite{elias57,elias58,elias88} that channels with a
zero-error capacity equal to $0$ can still transmit information
without any error, in the following sense: He introduced the
\emph{zero-error list-of-$L$ capacity} (with and without feedback), as
a generalization of Shannon`s zero-error capacity.  It defines the
asymptotically maximal rate at which bits can be transmitted by a
channel without any error, if the decoder is allowed to output a list
of $L$ values, where one of them must be the value sent by the sender.
For every channel that is non-trivial (one that cannot be simulated
without any underlying communication), there exists a value $L$ for
which the zero-error list-of-$L$ capacity is non-zero.  He also gave a
lower and upper bound for the zero-error list-of-$L$ capacity and
showed that it approaches the zero-error list-of-$L$ capacity with
feedback when $L$ increases.  The problem of optimal list-decodable
transmission has been further investigated in
\cite{frekos84,korner86,kormar88,kormar90}.

In this paper, we will take a slightly different perspective on
zero-error communication.  We will use a definition of channels
without probabilities: A channel only defines for every input symbol a
set of possible output symbols. We show that the smallest value of $L$
for which the zero-error list-of-$L$ capacity is non-zero completely
characterizes the possibility in principle of a channel to simulate
any other channel\footnote{Note that this question is trivial for
  ordinary channels and ordinary reductions: any non-trivial channel
  can simulate any other channel with a small error probability.}.  We
will call this value the \emph{ambiguity} of a channel, since it
characterizes the least ambiguity the receiver has over a value sent
by the sender.

In the second part of the paper, we will show how the ambiguity of a
\emph{network} of channels can be calculated, given the ambiguities of
each channel and a structure that defines which sets of channels may
be corrupted by malicious players.

\section{Definitions}

A channel is a conditional probability distribution that defines for
every input symbol the probability distribution of the output symbols.
However, since we are only interested in zero-error transmission, we will
use a simplified version of channels without probabilities, which we
will call \emph{zero-error channels}.  They only define which outputs
are \emph{possible}, but not how \emph{probable} they are.  While
still preserving all the characteristics of a channel needed in our
context, this definition has the advantage that we can not only use it
as a model for the communication primitive present, but also for the
communication that we try to achieve.  Furthermore, it can also be
applied in a context where the probabilities are not known or do not
exist. For example, one can think of situations where an malicious
player tries to manipulate the communication. He may always choose the
worst outcome for the receiver, knowing the protocol of the sender and
the receiver.
  
\begin{definition}
  A \emph{\Channel{\mX}{\mY}} is a relation $\mW \subseteq \mX \times
  \mY$, where $\mX$ is the input domain, $\mY$ the output range and
  $\mW$ the set of all possible input/output pairs. For every input
  symbol, there must exist at least one output symbol. $\forall x \in
  \mX: \{ y \in \mY \big | (x,y) \in \mW\} \neq \emptyset$.
\end{definition}

For simplicity, we will also use 
$\mW(x) = \{ y \in \mY
\big | (x,y) \in \mW\}$ to denote the set of valid output symbols for
the input symbol $x$.

\begin{definition}
  A $(\mX,\mY)$-\emph{protocol} using a \Channel{\mX_0}{\mY_0} $\mW$
  as communication primitive is an algorithm executed by the sender
  and the receiver, where the sender has an input $x \in \mX$ and the
  receiver an output $y \in \mY$ and the sender can send messages over
  $\mW$ to the receiver.
  
  In an $(\mX,\mY)$-\emph{feedback-protocol}, the receiver is
  additionally allowed to send values to the sender over a perfect
  channel.
\end{definition}

\begin{definition}
  Let $\mW_0$ be a \Channel{\mX_0}{\mY_0} and $\mW_1$ a
  \Channel{\mX_1}{\mY_1}. $\mW_1$ is \emph{achievable} by $\mW_0$
  ($\mW_0 \achieves \mW_1$) if there exists a
  $(\mX_1,\mY_1)$-\emph{protocol} using $\mW_0$ as communication
  primitive, such that for every input and for every possible output
  of the channel invocations, the receiver gets an output from the
  protocol that fulfills the requirements of $\mW_1$.
  
  If there exists $(\mX_1,\mY_1)$-\emph{feedback-protocol}, we say
  that $\mW_1$ is \emph{achievable with feedback} by $\mW_0$ ($\mW_0
  \achievesF \mW_1$).
\end{definition}

\section{Reduction of Channels} \label{List}

In this section, we introduce a special class of channels, the
\emph{List-channels}. We show that they are completely ordered with
respect to achievability and that every channel is equivalent to a
List-channel. Hence, \emph{all} channels are completely ordered with
respect to achievability.  The List-channels model the communication
achieved using list-decodable codes.

\begin{definition}
  Let $a, d \in \mathbb N$ with $a < d$. Let $\mX =
  \{1,\ldots,d\}$ and $\mY = \{y \subset \mX \big | |y| = a \}$.
  A $\Listd{a}{d}$-channel is a $(\mX, \mY)$-zero-error channel, with
  \[ \Listd{a}{d} = \{(x,y) \in \mX \times \mY \big | x \in y\}.\]
\end{definition}

We use $\List{a}$ as a short form of $\Listd{a}{a+1}$. $\List{\infty}$
denotes the trivial List-channel over which no communication is
possible.

In \cite{elias88}, it was proven in Proposition 2a that if all
$(L+1)$-tuples of input symbols of a channel are adjacent\footnote{ A
  tuple is adjacent if the output sets of all input symbols in the
  tuple intersect.}, then the list-of-$L$ feedback capacity of that
channel is~$0$. In fact, it is easy to see from the proof of
Proposition 2a that such a channel cannot simulate \emph{any}
$\Listd{a}{d}$ channel for $a \leq L$, even with feedback.

\begin{lemma} \label{CNot2List}
  Let $\mW$ be a \Channel{\mX}{\mY} and $L \in \mathbb N$. If
  \[ 
     \forall x_1,\ldots,x_{L+1} \in \mX:\mW(x_1) \cap \cdots 
     \cap \mW(x_{L+1}) \neq \emptyset,
  \]
  then, for all $d \in \mathbb N$ and $a \leq L$, we have that $\mW
  \nachievesF \Listd{a}{d}$.
\end{lemma}

However, Proposition 4 in \cite{elias88} states that if any
$(L+1)$-tuple of input symbols in a channel is not adjacent, then the
list-of-$L$ capacity (with and without feedback) is positive. This
means that such a channel can simulate all $\Listd{a}{d}$ channels
with $a \geq L$.

\begin{lemma} \label{C2List}
  Let $\mW$ be a \Channel{\mX}{\mY} and $L \in \mathbb N$. If
  \[ 
  \exists x_1,\ldots,x_{L+1} \in \mX:\mW(x_1) \cap \cdots \cap
  \mW(x_{L+1}) = \emptyset,
  \]
  then for all $d \in \mathbb N$ and $a \geq L$, we have that $\mW
  \achieves \Listd{a}{d}$.
\end{lemma}

We see that feedback does never increase the set of possible
List-channels that a channel can achieve.  From these two lemmas
follows now directly the following corollary, which states that the
List-channels are completely ordered with respect to achievability,
and that there exist infinite many equivalence classes.
Note that for all $\Listd{a}{d}$ channels, all
$a$-tuples of input symbols are adjacent, but none of the
$(a+1)$-tuples.

\begin{corollary} \label{S2S}
  For all $a$, $d$, $a'$ and $d'$, $\Listd{a}{d} 
  \achievesF \Listd{a'}{d'}$ holds exactly when
  $\Listd{a}{d} \achieves \Listd{a'}{d'}$ holds, namely if and only if
  $a \leq a'$.
\end{corollary}

We will now show that every channel is equivalent to a $\List{a}$
channel for a specific $a$. This means that all channel can in fact
be interpreted as a List-Channel.

\begin{theorem} \label{main}
  For every zero-error channel $\mW$ there exists exactly one $a \in
  \mathbb N \cup \{\infty\}$, such that $\mW \achieves \List{a}$ and
  $\List{a} \achieves \mW$. This value $a$ is called the
  \emph{ambiguity} of $\mW$, denoted by $A(\mW)$.
\end{theorem}

\begin{proof}
  If the output sets of all inputs intersect, the channel is trivial
  and therefore equivalent to the $\List{\infty}$ channel.
  Otherwise, let $a$ be the biggest number for which it is true that 
  \begin{eqnarray}
     \forall x_1,\ldots,x_{a} \in \mX:\mW(x_1) \cap \cdots 
     \cap \mW(x_{a}) \neq \emptyset. \label{condition}
  \end{eqnarray}
  From Lemma \ref{C2List} follows directly that $\mW \achieves \List{a}$.
  It remains to be shown that $\Listd{a}{|\mX|} \achieves \mW$, since 
$\List{a} \achieves \Listd{a}{|\mX|}$.
  Let $f: \mX \rightarrow \{1,\ldots,|\mX|\}$ be a bijective function. On
  input $x \in \mX$, the sender sends $f(x)$ over the channel. The
  receiver gets the values $v_1,\ldots,v_a$ and outputs
  \[y \in \mW(f^{-1}(v_1))\cap \cdots \cap \mW(f^{-1}(v_a)).\]
  Such a $y$ exists due to the Condition \ref{condition}.
\end{proof}

\begin{corollary}
  For all $\mW_1$ and $\mW_2$, 
$\mW_1 \achievesF \mW_2$ holds exactly when
$\mW_1 \achieves \mW_2$ holds, namely if and only if 
$A(\mW_1) \leq A(\mW_2)$.
\end{corollary}

The value $A(\mW)$ is therefore a measure for the possibility of
simulating other channels by the channel $\mW$ (if efficiency is of no
importance).  Since feedback never helps to increase the set of
achievable channels, it is sufficient to look at protocols without
feedback.

\section{Networks of Channels}

A message can also be indirectly sent to the receiver, through other
players. In this section we will show how the ambiguity of such a
communication can be calculated.

\begin{lemma} \label{2serie}
  Let $A$, $B$, and $C$ be three players, and let $\mW_1$ be a zero-error channel
  from $A$ to $B$ and $\mW_2$ a zero-error channel from $B$ to $C$.  The zero-error channel
  $\mW_s$ between $A$ and $C$ resulting from serial concatenation of $\mW_1$ and $\mW_2$ 
  has an ambiguity of $A(\mW_s) = A(\mW_1)A(\mW_2)$.
\end{lemma}

\begin{proof}
  $P_2$ can send all the values he received to $P_3$, which are at
  most $A(\mW_1)$. For each of these values, $P_3$ receives at most
  $A(\mW_2)$ values. Therefore we have $A(\mW_s) \leq
  A(\mW_1)A(\mW_2)$.
  
  On the other hand, the channels $\mW_1$ and $\mW_2$ can simulate
  $A(\mW_2)-1$ players $B_i$ and $A(\mW_1)A(\mW_2) - 1$ players $A_i$,
  such that all of the players $B_i$ receive messages from $A(\mW_1)$
  different senders.  Therefore we have $A(\mW_s) \geq
  A(\mW_1)A(\mW_2)$.
\end{proof}

\begin{corollary}
  Let $W =\{ \mW_1, \dots, \mW_n \}$ be a set of zero-error channels
and let $\mW_s$ be the serial concatenation if its elements.
  We have \[ A(\mW_s) = \prod_{i=1\dots n}A(\mW_i). \]
\end{corollary}

If \emph{any} of the intermediate players in a serial concatenation is
\emph{malicious} (that is a player who does not follow the protocol),
no communication is possible.  We say that the resulting channel is
malicious. Note that the adversary has now two different ways of
disturbing the communication: First of all, he controls the malicious
channels completely, and secondly he can choose all the additional
values the receiver gets over all the non-malicious channels.

Any transmission protocol for a network of players can be changed in
such a way that every intermediate players send the values to the next
player without any processing. All the processing is done by the
receiver. In any graph there exists a finite amount of paths without
cycles from the sender to the receiver. It is now easy to see that any
graph is equivalent to a parallel concatenation of channels,
which are build by serial concatenation of all the channels on a path.
All channels that have at least one intermediate malicious player are
malicious.  We will call the set of all the channels which are not
malicious the \emph{honest set}. Generally, it is not possible for the
receiver to know which channel belongs to the honest set and which do
not. However we can assume that he knows that the honest set is an
element of a \emph{honest set structure} $\mH$, which is the set of the
possible honest sets. Honest set structures are equivalent to the
general adversary structures, introduced in~\cite{hirmau97}.

The following theorem shows how the ambiguity of a parallel
concatenation of channels with a given honest set structure can be
calculated. Using the transformation above, it can be used to
calculate the ambiguity of a virtual channel between two players in
any network of players connected by channels.

\begin{theorem} \label{para1}
  Let $W =\{ \mW_1, \ldots, \mW_n \}$ be a set of zero-error channels and $\mH = \{h_1,
  \dots, h_k\}$ a honest set structure. Let $\mW_p$ be the parallel
  concatenation of the channels in $W$ with the honest set
  structure $\mH$. Let $S_j = \{i|j \in h_i \}$ be the set of
  indexes of all honest sets wherein player $j$ is. The ambiguity of
  $\mW_p$ is the maximum of the sum of some integers $a_1,\dots,a_k$,
  \[ A(\mW_p) = \max_{a_1,\dots,a_k}{\sum_{i=1}^k{a_i}} \]
  such that for all $j \in \{1,\dots,n\}$
  \begin{eqnarray}
   \sum_{i \in S_j}{a_i} \leq A(\mW_j) \label{a-cond}
  \end{eqnarray}
  holds.
\end{theorem}

\begin{proof}
  First of all, we show that there exists a strategy by the sender and
  the receiver to transmit a value with an ambiguity of at most
  $A(\mW)$. The sender sends his value through all channels. The
  receiver takes all the values for which there exists a honest set
  such that all of the channels in that honest set output that value.
  Assume that the receiver outputs for the honest set $h_i$ $b_i$
  values. Because every channel $\mW_j$ can output at most $A(\mW_j)$
  values, we have for all channels $\mW_j$ that
  \[ \sum_{i \in S_j}{b_i} \leq A(\mW_j). \]
  The receiver outputs $\sum_{i=1\dots k}{b_i}$ values, which is not
  bigger than $A(\mW_p)$ since $A(\mW_p)$ maximizes this sum.
  
  It remains to be shown that there exists a strategy by the adversary
  such that $\mW_p$ has an ambiguity of at least $A(\mW_p)$.  The
  adversary simulates $A(\mW_p)-1$ senders. He chooses a honest set
  $h_r$ with $a_r > 0$ and corrupts all channels not in $h_r$.  He
  distributes the $A(\mW_p)-1$ other senders among all honest sets,
  such that every honest set $h_i$ gets $a_i$ senders. Every channels
  outputs all the values from all the honest sets it belongs to, which
  is possible for all channels $\mW_j$ due to the Equations
  \ref{a-cond}.
  Since for the receiver any of the sender could be the real sender,
  the ambiguity $\mW_p$ is at least $A(\mW_p)$.
\end{proof}

$A(\mW)$ can be calculated using linear programming. But because the
structure $\mH$ can be very big, it may take a lot of time to
calculate it.
However, if we only have a threshold honest set structure, that is, if up to
$t$ channels are malicious, then the ambiguity of the parallel
concatenation is much easier to calculate.

\begin{theorem} \label{threshold}
  Let $W =\{\mW_1, ...,\mW_n\}$ be a set of channels, from which $t$
  channels may be malicious. Then the ambiguity of the parallel
  concatenation $\mW_p$ of the channels in $W$ is
  \[A(\mW_p) = \min_{G \subseteq W} \left \lfloor \frac{\sum_{\mW_i 
  \in G}{A(\mW_i)}}{|G| - t} \right \rfloor.\]
\end{theorem}

\begin{proof}
  Again, we show that there is a strategy by the sender and the
  receiver to get an ambiguity which is at most $A(\mW_p)$.  The
  sender sends his value through all channels. The receiver takes all
  the values he gets from the channels in $G$, where $G$ is a set for
  which it holds that
  \[\left \lfloor \frac{\sum_{\mW_i \in G}{A(\mW_i)}}{|G| - t} \right 
  \rfloor = A(\mW_p).\]
  He outputs all values that occur at least $|G| - t$ times. Because
  at most $t$ channels are malicious, at least $|G| - t$ channels will
  output the value sent by the sender and therefore the correct value
  will be output by the receiver.  Furthermore, he outputs at most
  $A(\mW_p)$ values.
  
  The adversary can use the following strategy to achieve an ambiguity
  of at least $A(\mW_p)$. He simulates $A(\mW_p)-1$ senders. On the
  channels with an ambiguity bigger than $A(\mW_p)$, he simply sends
  all $A(\mW_p)$ values. Let $\hat G$ be the set of all the channels
  with an ambiguity smaller than $A(\mW_p)$.  The adversary corrupts
  $t$ channels in $\hat G$ and sends the output of all the $A(\mW_p)$
  senders $|\hat G| - t$ times, distributed over all the channels in
  $\hat G$ such that all values sent by one channel are different.
  This is possible since none of them has an ambiguity bigger than
  $A(\mW_p)$ and since
  \[A(\mW_p) \leq \frac{\sum_{\mW_i \in \hat G}{A(\mW_i)}}{|\hat G| - t}. \]
  The receiver cannot know which of the $A(\mW_p)$ senders is the real
  sender.
\end{proof}

This optimum can be found efficiently by sorting the channels
according to their ambiguity. The following corollaries follow
directly from Theorem~\ref{threshold}.

\begin{corollary}
  Let $W =\{\mW_1, ...,\mW_n\}$ be a set of channels with the same
  ambiguity $A(\mW_1) = \dots = A(\mW_n) = A$ and from which $t$
  channels may be corrupted. Then the ambiguity $A(\mW_p)$ of the the
  parallel concatenation $\mW_p$ of the channels $\mW_1, ...,\mW_n$ is
  \[A(\mW_p) = \left \lfloor \frac{n}{n - t} A\right \rfloor. \]
\end{corollary}

\begin{corollary}
  Let $W =\{\mW_1, ...,\mW_n\}$ be a set of channels and let none of
  them be corrupted ($t=0$).  Then the ambiguity $A(\mW_p)$ of the
  parallel concatenation $\mW_p$ of the channels in $W$ is
  \[A(\mW_p) = \min(A(\mW_1), \ldots, A(\mW_n)).\]
\end{corollary}

The following corollary restates a result from \cite{dolev82}, namely
that a majority of honest $\List{1}$ channels is needed in order to
simulate a $\List{1}$ channel.

\begin{corollary}
  A $\List{1}$ channel can be simulated by $n$ $\List{1}$ channels, from which
  up to $t$ may be corrupted, if and only if $n > 2t$.
\end{corollary}

\end{document}